\documentclass[nature,letterpaper,twocolumn,superscriptaddress,amsmath,amsfonts,amssymb,preprintnumbers,citeautoscript]
{revtex4}
\pdfoutput=1
\usepackage[T1]{fontenc}\usepackage[latin1]{inputenc}
\usepackage{dcolumn,graphicx,color,booktabs} \graphicspath{{Images/}}
\usepackage{amsmath,bm}
\renewcommand{\figurename}{Fig.}
\makeatletter\renewcommand{\fnum@figure}[1]{\figurename\thefigure.}\makeatother
\definecolor{DarkBlue}{rgb}{0,0,0.5}

\begin{document} \pagestyle{plain}

\title{Electronic band structure and momentum dependence of the superconducting gap in Ca$_{1-x}$Na$_x$Fe$_2$As$_2$ from angle-resolved photoemission spectroscopy}

\author{D.\,V.\,Evtushinsky}
\affiliation{Institute for Solid State Research, IFW Dresden, P.\,O.\,Box 270116, D-01171 Dresden, Germany}
\author{V.\,B.\,Zabolotnyy}\author{L.\,Harnagea}
\affiliation{Institute for Solid State Research, IFW Dresden, P.\,O.\,Box 270116, D-01171 Dresden, Germany}
\author{A.\,N.~Yaresko}
\affiliation{Max-Planck-Institute for Solid State Research, Heisenbergstrasse 1, D-70569 Stuttgart, Germany}
\author{S.\,Thirupathaiah}
\affiliation{Institute for Solid State Research, IFW Dresden, P.\,O.\,Box 270116, D-01171 Dresden, Germany}
\author{A.\,A.\,Kordyuk}
\affiliation{Institute for Solid State Research, IFW Dresden, P.\,O.\,Box 270116, D-01171 Dresden, Germany}
\affiliation{Institute of Metal Physics of National Academy of Sciences of Ukraine, 03142 Kyiv, Ukraine}
\author{J.\,Maletz}\author{S.\,Aswartham}
\affiliation{Institute for Solid State Research, IFW Dresden, P.\,O.\,Box 270116, D-01171 Dresden, Germany}
\author{S.\,Wurmehl}
\affiliation{Institute for Solid State Research, IFW Dresden, P.\,O.\,Box 270116, D-01171 Dresden, Germany}
\affiliation{Institut f\"{u}r Festk\"{o}rperphysik, Technische Universit\"{a}t Dresden, D-01171 Dresden, Germany}
\author{E.\,Rienks}\author{R.\,Follath}
\affiliation{BESSY GmbH, Albert-Einstein-Strasse 15, 12489 Berlin, Germany}
\author{B.\,B\"{u}chner}
\affiliation{Institute for Solid State Research, IFW Dresden, P.\,O.\,Box 270116, D-01171 Dresden, Germany}
\affiliation{Institut f\"{u}r Festk\"{o}rperphysik, Technische Universit\"{a}t Dresden, D-01171 Dresden, Germany}
\author{S.\,V.\,Borisenko}
\affiliation{Institute for Solid State Research, IFW Dresden, P.\,O.\,Box 270116, D-01171 Dresden, Germany}

\begin{abstract}
\noindent Electronic structure of newly synthesized single crystals of calcium iron arsenide doped with sodium with $T_{\rm c}$ ranging from 33 to 14\,K has been determined by angle-resolved photoemission spectroscopy (ARPES). The measured band dispersion is in general agreement with theoretical calculations, nonetheless implies absence of Fermi surface nesting at antiferromagnetic vector. A clearly developing below\,$T_{\rm c}$ strongly band-dependant superconducting gap has been revealed for samples with various doping levels. BCS ratio for optimal doping, $2\Delta/k_{\rm B}T_{\rm c}=5.5$, is substantially smaller than the numbers reported for related compounds, implying a non-trivial relation between electronic dispersion and superconducting gap in iron arsenides.
\end{abstract}


\maketitle

Iron-based high temperature superconductors form an increasingly growing subject for investigation. Unlike other types of high-temperature superconductors, iron-based compounds can be synthesized in form of various crystals, exhibiting a large variety of electronic band structures, magnetic properties, superconducting order parameters, and electronic properties in general \cite{Paglione, Stewart, Kord}. On the other hand, experimental difficulties, connected to the limitations of each particular technique, and complexity of the electronic interactions in iron-based materials hinder accurate and precise determination of the electronic structure in many cases. Thus, despite such abundance of possible forms of iron-based superconductors, the electronic spectrum, including both electronic band structure and superconducting gap function, has been thoroughly addressed experimentally only for hole-doped BaFe$_2$As$_2$ \cite{DingEPL, FengPRL, VolodyaNature, EvtushinskyPRB, BorisPRL, Orbitalgap} and LiFeAs \cite{BorisenkoLiFeAs, Sergey_LiFeAs1, Ding_LiFeAs}. Available information at hand is still insufficient to tell, which features are generic to all iron-based high-$T_{\rm c}$s and responsible for the effective electron pairing.

\begin{figure}[]
\includegraphics[width=1\columnwidth]{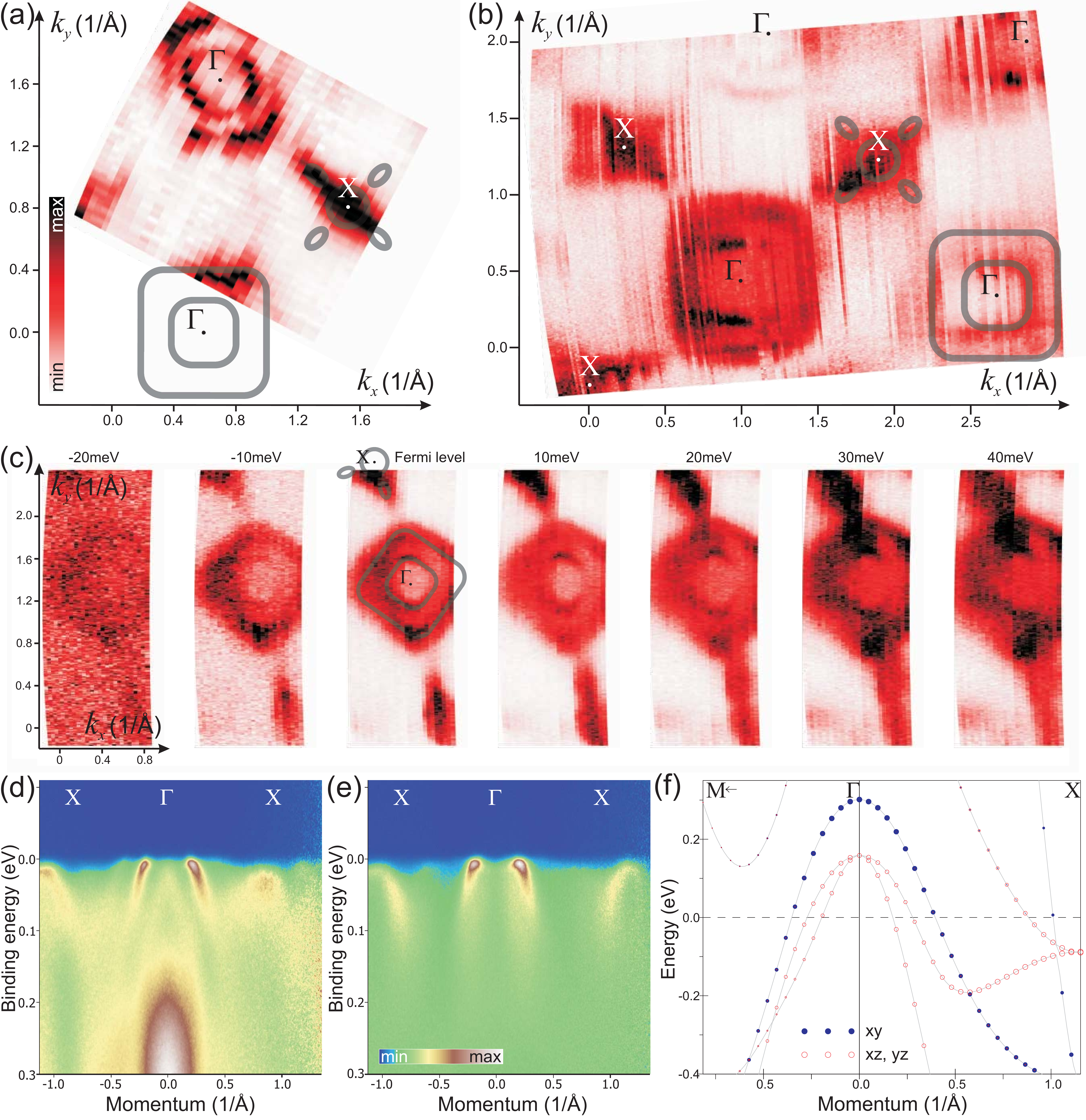}
\vspace{-0.3cm}
\caption{(a,b) Fermi surface maps of Ca$_{1-x}$Na$_x$Fe$_2$As$_2$ with $x=0.68\pm0.04$ and $T_{\rm c}=33$\,K, recorded at 40 and 80\,eV excitation energy ($h\nu$) respectively. (c) Cuts through photoemission intensity distribution at different binding energies reveal hole-like nature of the propellers blades, recorded at $h\nu=80$\,eV. Energy-momentum cut passing close to $\Gamma$X line recorded with 90\,eV-horizontally (d) and -vertically (e) polarized light. (f) Theoretical band dispersion.}
\vspace{-0.3cm}
\label{FS_maps}
\end{figure}

\begin{figure*}[]
\includegraphics[width=0.95\textwidth]{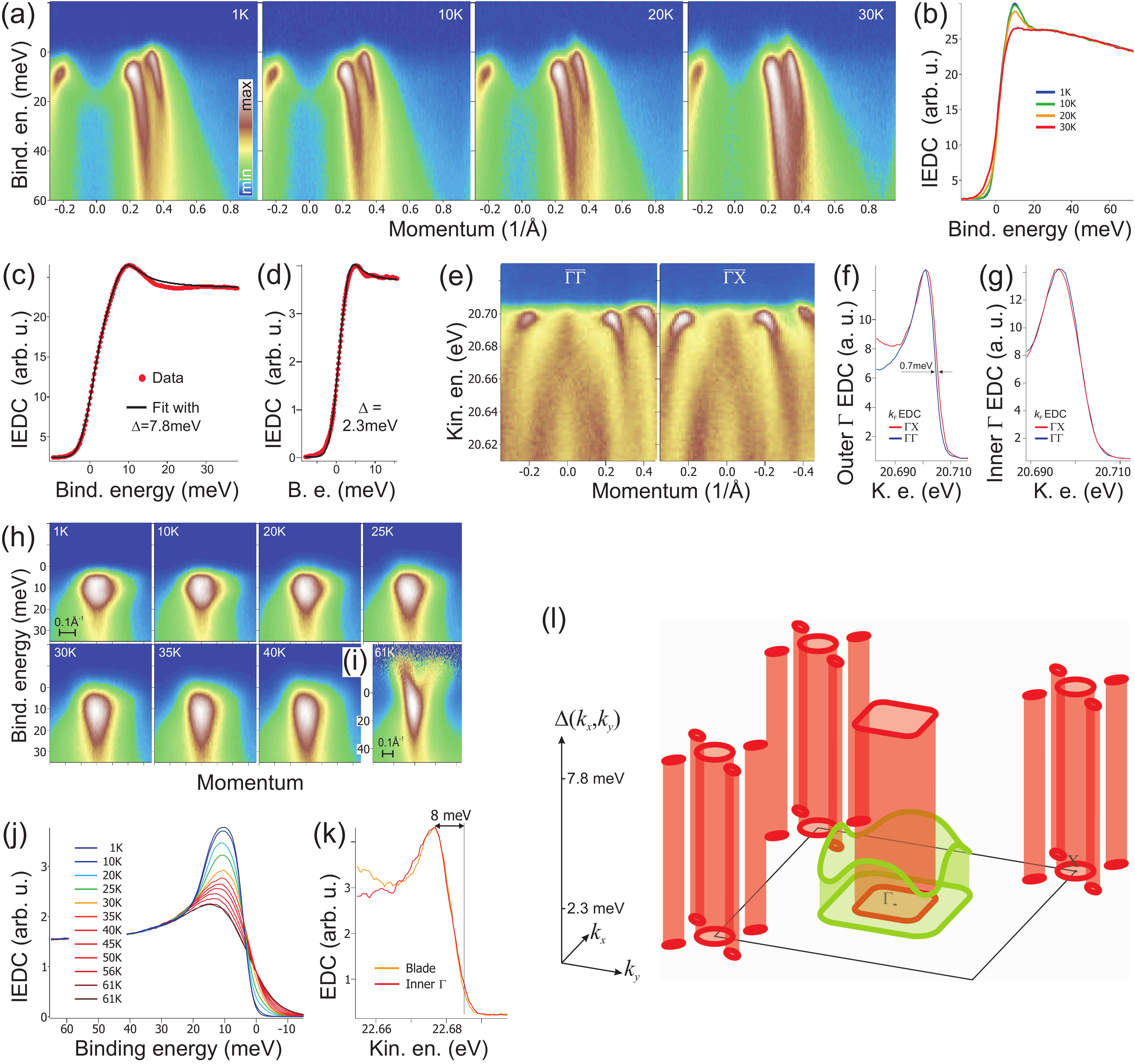}
\caption{(a) Temperature dependence of the energy-momentum cut passing through the $\Gamma$ point for the sample with $T_{\rm c}=33$\,K. (b) Temperature dependence of the partial density of states, i.e. integrated energy distribution curves (IEDC). (c) Gap size, extracted from fit IEDC to Dynes function, equals 7.8\,meV for the two $xz/yz$ inner $\Gamma$ barrels. (d) A similar analysis for the outer $\Gamma$ barrel results in the gap magnitude of 2.3\,meV. (e) Energy-momentum cuts along $\Gamma\Gamma$ and $\Gamma$X. (f,g) $k_{\rm F}$ EDC for the outer and inner $\Gamma$ barrels at $\Gamma\Gamma$ and $\Gamma$X. (h) Temperature dependence of the X pocket. (i) 61\,K-spectrum divided by Fermi function. (j) Temperature dependence of IEDC from X pocket. (k) $k_{\rm F}$ EDC for inner $\Gamma$ barrel and propeller's blade. (l) Momentum dependence of the superconducting gap in Ca$_{1-x}$Na$_x$Fe$_2$As$_2$.}
\label{Gap_Gamma}
\end{figure*}

Here we report detailed studies of the electronic structure of newly synthesized large single crystals of Ca$_{1-x}$Na$_x$Fe$_2$As$_2$ (CNFA), with doping level, $x$, up to $0.7$ and $T_{\rm c}$ between 14 and 33\,K. Angle-resolved photoemission spectroscopy (ARPES) measurements were performed at $1^3$ end station at BESSY II synchrotron in Berlin \cite{OneCube, JOVE}. The sample surface was prepared by cleaving, and was shown to be highly suitable for ARPES experiments: sharp spectral features and a pronounced superconducting transition at nominal $T_{\rm c}$ were observed, offering possibility for detailed studies of the electronic band structure and superconducting gap distribution with high resolution.



The Fermi surface (FS) of CNFA consists of a propeller-shaped structure at the Brillouin zone (BZ) corner, alike other hole-doped 122 compounds, and hole-like FS sheets at the BZ center of increased size [see Fig.~1(a,b)]. Figure~1(c) shows cuts through the photoemission intensity distribution at different binding energies, and the hole-like nature of the propeller's blades is confirmed by the increase of their size with binding energy. Figures~1(d) and (e) show an energy-momentum cut, passing close to the high-symmetry $\Gamma$X line. Dispersion of the hole-like bands supporting $\Gamma$ barrels and propeller blades is seen. Panel (f) shows the band dispersion, obtained for CaFe$_2$As$_2$ theoretically. Band structure calculations were performed using the LMTO method for the experimental crystal structure of Ca$_{1-x}$Na$_{x}$Fe$_2$As$_2$ with $x$=0.5 \cite{zlwd11}. The effect of doping by 0.5 hole per Fe on the bands was accounted for by a rigid band shift.

The revealed electronic band structure of CNFA is generally similar to the electronic structure of the well studied Ba$_{1-x}$K$_x$Fe$_2$As$_2$ (BKFA). The band dispersion of CNFA differs from the band dispersion of BKFA in the following way: (i) the sizes of the hole-like $\Gamma$ barrels are increased, in accordance with higher level of hole doping; (ii) both outer and inner $\Gamma$ barrels are more squarish; (iii) splitting between the outer and inner $\Gamma$ barrels is larger; (iv) the electron pocket is larger, and the propeller blades are smaller; (v) the electronic states at $(k_x=0, k_y=0)$, presumably originating from $3z^2-1$ orbitals \cite{Orbitalgap, Yoshida_KFA}, are located closer to the Fermi level. Additionally a seeming splitting of the outer $\Gamma$ barrel is observed for CNFA, while no such effect has been observed in the spectra of BKFA, where the outer $\Gamma$ barrel appears as the most sharp feature. The latter discrepancy between CNFA and BKFA spectra is perfectly explained by the band structure calculations: although calculated band structures of CaFe$_2$As$_2$ (CFA) and BaFe$_2$As$_2$ (BFA) are very similar, there are important differences caused by stronger interlayer coupling in CFA \cite{ab11}. In particular, the top of a Fe $d_{3z^2-1}$ band at the Z point, which in BFA lies below the Fermi level \cite{ylaa09}, shifts above the top of a two dimensional $d_{xy}$ band. As a consequence, at Z point the outmost hole-like FS is formed by the strongly dispersing $d_{3z^2-1}$ band. Thus comparison to theoretical band structure allows us to identify inner $\Gamma$ barrel as originating from $d_{xz/yz}$ orbitals of iron, outer $\Gamma$ barrel\,---\,as $d_{xy}$, and propeller as mainly $d_{xz/yz}$ (see also Ref.\,\onlinecite{Orbitalgap}). Without in-depth analysis we estimate band renormalization for CNFA as 2.5--3, which is close to other iron arsenides \cite{DingEPL, BorisPRL, Orbitalgap}.

Figure~2 shows temperature-dependent measurements, revealing opening of the superconducting gap in the photoemission spectra. The energy-momentum cut, shown in panel (a), captures Fermi crossings of the inner and outer $\Gamma$ barrels. Bending of the band dispersion is seen for both outer and inner $\Gamma$ barrels with stronger effect for the latter. Figure~2(b) shows temperature dependence of the partial density of states, exhibiting growth of the coherence peak below $T_{\rm c}$. Partial density of states for the inner band measured at 1\,K is shown in the panel (c) with a fit to the Dynes function \cite{Dynes, EvtushinskyPRB}. Fitting yields a gap value of 7.8\,meV. Corresponding analysis for the outer band yields gap magnitude of 2.3\,meV (d). Noticeable non-superconducting component is present in the photoemission spectra, similar to the case of the  hole-doped BaFe$_2$As$_2$ \cite{DingEPL, EvtushinskyPRB, Orbitalgap}.

Figures~2(e) shows energy--momentum cuts recorded at $h\nu=25$\,eV along $\Gamma$X and $\Gamma\Gamma$ directions. While there is virtually no difference between these directions for the gap on the inner $\Gamma$ barrel, the back-bending dispersion, which is a signature of a gap, at the outer $\Gamma$ barrel is substantially more pronounced for $\Gamma \Gamma$ direction. Further analysis of $k_{\rm F}$ energy distribution curves (EDCs) [see panels (f, g)] shows that spectrum for $\Gamma \Gamma$ direction is shifted further from the Fermi level, as compared to $\Gamma$X, suggesting that superconducting gap at the outer $\Gamma$ barrel is anisotropic with minima along the Brillouin zone diagonal, $\Gamma$X line. Though this anisotropy is rather weak and generally is within error bars of experiment, several data sets indicate that it is as described above: for the outer $\Gamma$ FS sheet the gap is smaller along $\Gamma$X and the modulation magnitude is comparable to the gap size, as opposed to the inner $\Gamma$ FS sheet, for which the modulation is much smaller than the gap size itself.

\begin{figure}[]
\includegraphics[width=0.96\columnwidth]{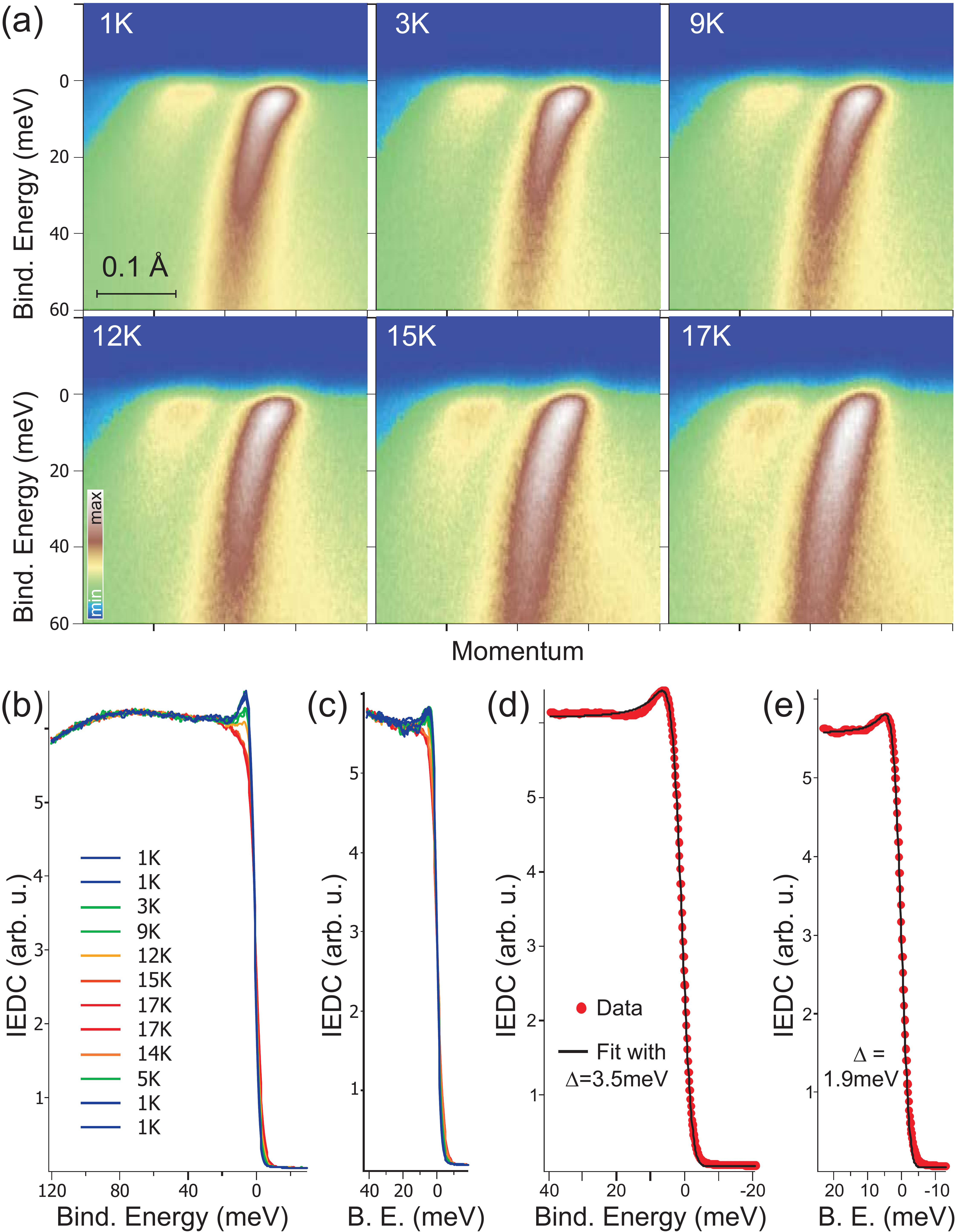}
\caption{(a) Temperature dependence of an energy-momentum cut passing through the $\Gamma$-point for Ca$_{1-x}$Na$_x$Fe$_2$As$_2$ sample with $T_{\rm c}=14$\,K. (b) Temperature dependence of the partial density of states for the inner $\Gamma$ barrel. (c) The same for the outer $\Gamma$ barrel. (d) and (e) data at 1\,K and fits to Dynes function for inner and outer $\Gamma$ barrels respectively. The derived gap sizes are 3.5 for the inner and 1.9\,meV  for the outer $\Gamma$ barrels.}
\label{Sample1_Tc15_Tdep}
\end{figure}

Figure~2(h) shows the temperature dependence of the electron-like X pocket. Flattening of the spectrum top below $T_{\rm c}$ indicates opening of the superconducting gap here. Panel (i) shows a high-temperature spectrum of X pocket divided by Fermi function, clearly confirming the electron-like nature of these FS sheet. Even better the onset of superconductivity is seen in Fig.2~(j), where the partial DOS for X pocket is presented: the peak grows below $T_{\rm c}$ and the leading edge shifts towards higher binding energies. Spectra, taken from the inner $\Gamma$ barrel and propeller blades at same conditions, indicate a uniform gap for the entire propeller structure [see panel (k)]. Additionally the electronic states of the mentioned above $3z^2-1$ band at $(k_x=0, k_y=0)$ exhibit a clearly noticeable response to the superconducting transition with a gap comparable to the one found for the outer $\Gamma$ barrel. The in-plane momentum dependence of the superconducting gap is summarized in the Fig.~2(l). The found in CNFA momentum distribution of the superconducting gap is very similar to the one that has been observed in BKFA \cite{EvtushinskyPRB, Orbitalgap}, except for the suggested large relative variation of the gap function for the outer $\Gamma$ barrel. Dependence of the gap on the out-of-plane electron momentum has been also measured, and is generally similar to those previously reported  for BKFA \cite{Orbitalgap} and BaFe$_2$As$_{2-2x}$P$_{2x}$ \cite{Feng_BFAP_node}: sharp minima in the gap as a function of $k_z$ occur at Z point, where $d_{3z^2-1}$ band interacts with $d_{xz/yz}$ bands of $\Gamma$ barrels.

\begin{figure}[]
\includegraphics[width=0.75\columnwidth]{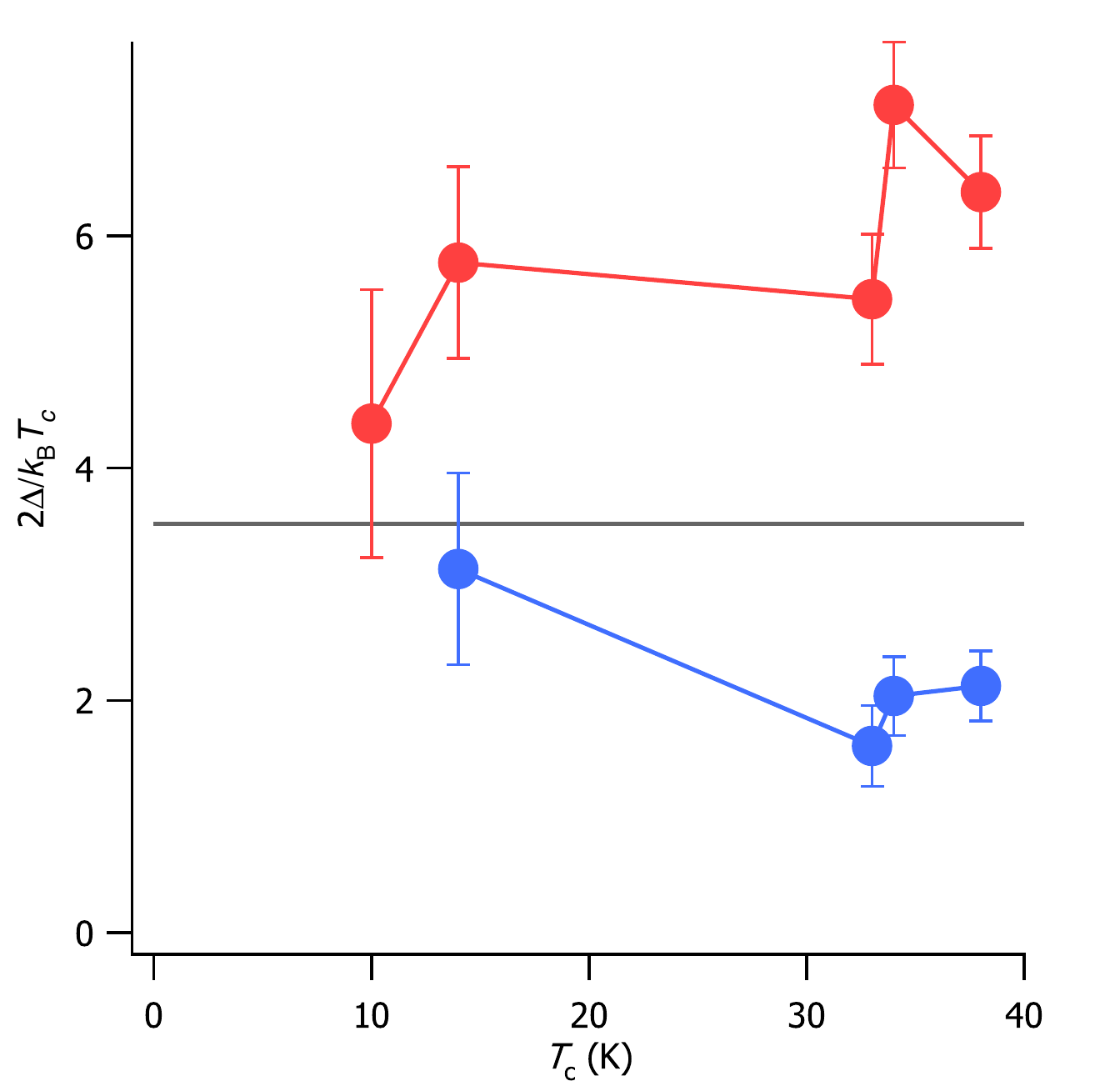}
\caption{BCS ratio $2\Delta/k_{\rm B}T_{\rm c}$ for several hole-doped 122 iron arsenides: optimally doped Ba$_{1-x}$K$_x$Fe$_2$As$_2$ with $T_{\rm c}=38$\,K \cite{Orbitalgap}, optimally doped ($T_{\rm c}=34$\,K) and underdoped ($T_{\rm c}=10$\,K) Ba$_{1-x}$Na$_x$Fe$_2$As$_2$ \cite{Sai}, optimally doped ($T_{\rm c}=33$\,K) and underdoped ($T_{\rm c}=14$\,K) Ca$_{1-x}$Na$_x$Fe$_2$As$_2$. Red dots represent maximal gap for the inner $\Gamma$ barrel, blue dots represent outer $\Gamma$ barrel. All values were extracted from ARPES data by fitting partial density of states to Dynes function.}
\label{Delta_Tc}
\end{figure}

Additionally we have succeeded to measure the superconducting gap opening for a Ca$_{1-x}$Na$_x$Fe$_2$As$_2$ phase with $T_{\rm c}$ of 14\,K. The temperature dependence of the energy-momentum cut passing through the BZ center is shown in Fig.~3(a). Figures~3(b) and (c) present a temperature dependence of the partial density of states for inner and outer $\Gamma$ barrels respectively. Figures~(d) and (e) show 1\,K-densities of states fitted with Dynes function. The derived values for the superconducting gap are equal 3.5 and 1.9\,meV for inner and outer $\Gamma$ barrel respectively.

In Fig.~4 we summarize the relation between the gap magnitude for different hole-doped 122 iron arsenides. All available compounds, Ba$_{1-x}$K$_x$Fe$_2$As$_2$, Ba$_{1-x}$Na$_x$Fe$_2$As$_2$, and  Ca$_{1-x}$Na$_x$Fe$_2$As$_2$ exhibit similar momentum dependence of the superconducting gap with drastic difference between the large gap (at inner $\Gamma$ barrel and propeller) and the small gap (at outer $\Gamma$ barrel), inherent also to many other iron arsenides \cite{EvtushinskyNJP}. One can clearly observe some common features for the behavior of these large and small gaps: $2\Delta/k_{\rm B}T_{\rm c}$ for the large gap is always more than the universal BCS value of 3.52, while for the smaller gap it is smaller; the deviation of the large gap upwards from the universal value is roughly proportional to the deviation of the small gap downwards. However, it is rather difficult to argue that $\Delta$ as a function of $T_{\rm c}$ can be approximated by a single smooth curve for all compounds, in particular there are Ca$_{1-x}$Na$_x$Fe$_2$As$_2$ and Ba$_{1-x}$Na$_x$Fe$_2$As$_2$ samples having virtually same $T_{\rm c}$ (33 and 34\,K) and substantially different gaps (7.8 and 10.5\,meV).
Merging of gap magnitudes to the BCS value for samples with low critical temperatures, also observed in specific heat studies of BaFe$_{2-x}$Co$_x$As$_2$ \cite{Hardy}, is consistent with decrease of electron-mediator coupling strength \cite{Carbotte}. Absence of strong universality in $\Delta(T_{\rm c})$ curve can stem from differences of electron pairing due to moderately varying Fermi surface geometry of the considered compounds. Alternatively the universality of the gap magnitude can be disturbed by obviously present here interplay of superconducting and magnetic orders \cite{Vavilov}.

In conclusion, we have pointed out a general similarity and particular dissimilarities of the Ca$_{1-x}$Na$_x$Fe$_2$As$_2$ electronic structure as compared to other 122 iron arsenide superconductors. Rather complicated Fermi surface with clear departures from the theoretical calculations is established as a characteristic property of the hole-doped 122 iron arsenides with highest $T_{\rm c}$s. Momentum dependence of the superconducting gap appears very similar for all studied compounds too, suggesting universality of the pairing mechanism.

We thank M.\,A.-H.\,Mohamed and S.-L.~Drechsler  for helpful discussions, and to R.\,H\"{u}bel for technical support. The work was supported under grants No. BO1912/2-2, BE1749/13 and WU595/3-1.


%
%
%
%
%
%
%
%
%
%

%
%
%
%
%
%
%
%
%
%
%
%
%
%
%
%
%

\end{document}